# RNA Binding Density on *X*-chromosome Differing from that on 22 Autosomes in Human

LV Zhanjun*, LU Ying**, SONG Shuxia*, ZHAI Yu*, WANG Xiufang *

*Department of Laboratory Animal, Hebei Medical University, Shijiazhuang 050017, China
**Center for Theoretic Biology Peking University, Peking 100871, China

**Abstract**
To test whether *X*-chromosome has unique genomic characteristics, *X*-chromosome and 22 autosomes were compared for RNA binding density. Nucleotide sequences on the chromosomes were divided into 50kb per segment that was recoded as a set of frequency values of 7-nucleotide (7nt) strings using all possible 7nt strings ($4^7=16384$). 120 genes highly expressed in tonsil germinal center B cells were selected for calculating 7nt string frequency values of all introns (RNAs). The binding density of DNA segments and RNAs was determined by the amount of complement sequences. It was shown for the first time that gene-poor and low gene expression *X*-chromosome had the lowest percentage of the DNA segments that can highly bind RNAs, whereas gene-rich and high gene expression chromosome *19* had the highest percentage of the segments. On the basis of these results, it is proposed that the nonrandom properties of distribution of RNA highly binding DNA segments on the chromosomes provide strong evidence that lack of RNA highly binding segments may be a cause of *X*-chromosome inactivation.

**Key works:** *X*-chromosome inactivation; intron RNA; DNA sequence composition; nucleotide string;

**Introduction**
Among placental mammals, the basic features of *X*-chromosome inactivation (XCI) are well established (BAILEY *et al.* 2000, OKAMOTO *et al.* 2000). XCI is the distinct difference between *X*-chromosome and autosomes. One of the two *X*-chromosomes of females becomes transcriptionally inactive in every cell of the early embryo and remains so in all somatic cells throughout life (COHEN and LEE 2002). This process serves to maintain the correct dosage relationship of genes between females (*XX*) and males (*XY*).
As for mechanism of the *X* inactivation, *Xist* has been shown by transgenic and knockout experiments in mice to play a pivotal role in initiating *X* inactivation (KELLEY and KUROD 2000). The 21.4 kb chicken lysozyme (cLys) chromatin domain was inserted into *X*-chromosome. The gene was expressed normally from the active *X*, but not resistant to XCI (CHONG *et al.* 2002). These works seem to provide evidences that *Xist* gene induces XCI.
If autosomal segments are attached to the *X* chromosome by translocation, XCI signal can spread for long distances into the autosome, (White et al 1998). However, spread into autosomal material extends less far and is less effective in gene silencing than in the *X* chromosome itself (White et al 1998, Rastan 1983, SOLARI *et al.* 2001). Thus, there is a barrier to the spreading effect of *X*-chromosome inactivation. The studies of gene expression in ICF female provide the examples of abnormal escape from *X*-chromosome inactivation in fibroblasts while *Xist* RNA localization is normal in these cells (HANSEN *et al.* 2000). Once XCI has been established, it is epigenetically inherited even if *Xist* is removed (BROWN and WILLARD 1994). These results argue against an independent silencing role for this RNA in somatic cells.
The majority of the genome is represented by repetitive sequences and nonprotein-coding sequences (Makalowski 2001). Why so many noncoding nucleotides? Zuckerkandl proposes that every nucleotide is at least infinitesimally functional (Zuckerkandl 2002). There are plenty of nonprotein-coding RNAs in cells (MATTICK 2001), the binding specificity of DNA-RNA is far higher than that of DNA-protein and the affinity of DNA with RNA is increased, as compared with DNA. It has been proved that RNA can open chromatin and increase it's DNase Ι sensitivity (ZHANG et al 2002, LV et al 2003), and triple chain nucleic acid can't form nucleosome (WESTIN et al 1995). Sensitizing to DNase I seems to be a general phenomenon of actively transcribed genes. These facts suggest that the interaction of RNA-DNA in cells may be extensive and functional.

In this paper to test whether *X*-chromosome has unique genomic characteristics we compared DNA sequence composition on *X*-chromosome to that on autosomes in term of RNA binding density that is a computer simulation of binding density of DNA segments and RNAs, and described nuclear sequences by frequencies of nucleotide strings that involve every nucleotide.

## Materials and methods

### Sequence data

Nucleotide sequences of 23 chromosomes were obtained from the NCBI (The version of the genome is "build 33", available on April 29.) (http://www.ncbi.nlm.nih.gov/genome/guide). Based on results of Digital Differential Display (DDD) (http://www.ncbi.nlm.nih.gov/UniGene.ddd.cgi), 120 genes highly expressed in tonsil germinal center B cells were selected. They are: *RPL13, YY1, GLTSCR2, KIAA0217, INPP5D, NCF1, NCUBE1, PTBP1, MLL, TCF3, BACH2,.YWHAQ, UBE2H, CGGBP1, CDC2, GGA2, SERP1, EGLN2, DUT, BCL2L12, WSB1, PTEN, MBNL, PAX5, BCL11A, FTH1, SMC4L1, CSNK1A1, OSBPL8, WHSC1, ALOX5, KIAA0084, ZFP91, KRAS2, FBP17, UGCG, ZNF265, FUSIP1, FOXP1, CYorf15B, CAMK2D, C9orf5, CLSTN1, DC8, CENTB2, NKTR, STK39, RERE, PSP1, PBP, MBP, DAPP1, FLJ11273, KIAA1323, NAP1L1, RASGRP1, CPNE3, UNC93B1, KIAA1033, ARS2, UBQLN1, LYN, TOMM20-PENDING, KIAA0746, PB1, MFNG, HSPCA, EIF4EBP2, GLS, OAZIN, FLJ22301 ,EHD1 ,ELF1, KIAA1268 ,FLJ10342 ,CEP1, BART1, BTF, FLJ20333, RCOR, GDI2, FLJ10407, APLP2, HNRPH1, MGC4796, CASP8, PTPRCAP, HRB2, PR KACB, MEF2B, NOLC1, LZ16, CAST, ADD3, AKAP13, AES, FLJ10392, FLJ20085, PSCD1, EIF2AK3, DDX18, CYBB, NAP1L4, PPP3CC, FLJ10707, CHERP, KIAA0494, DMTF1, RER1, MYBL1, FANCA, HSD17B12, CBX3, GNAS, NUP153, RANBP2, JJAZ1, ATM, ICAP-1A* and *NUP88*

### Software

Search software used for analysis of frequency values of 7-nucleotide strings was written by Lu Ying in our team. The 50kb DNA sequence was represented by a long column of numbers whose sum was 49994. MS EXCEL software was used for statistical analysis.

### Sequence representation

DNA sequences on chromosomes were divided into many segments of 50kb that were recoded for frequency values of all possible 7nt ($4^7$=16384). The incomplete segments in which nucleotides have been known more than 10% were selected for the count (using adjusting frequency values) and those less than 10% were neglected. The counting method for adjusted frequency values of incomplete segments is as follows:

$$P_i^k = \frac{49994}{\sum_{k=1}^{16384} p_i^k} * p_i^k$$

Where $P_i^k$ is the adjusted frequency value of the $k^{th}$ 7nt in segment $i$, $p_i^k$ is its original value, and 49994 is the sum of frequency values of 50kb DNA segment.

120 genes highly expressed in tonsil germinal center B cells were selected for calculating 7nt string frequency values of all introns (from sense strand). Each intron sequence was recoded initially as a set of 7nt string frequency values. The sum of 7nt string frequency values on same 7nt string of all introns within same gene multiplied by expressing frequency of the gene was intron 7nt string frequency values of this gene. The sum of intron 7nt string frequency values of 120 genes (intron 7nt) was regarded as a simulation of RNA fragments (RNAs) in cells.

**RNA binding density to DNA sequences**
After knowing the 7nt string frequency values of DNA segments and intron 7nt, we can calculate the binding density of DNA to intron 7nt. Intron 7nt is RNA except T substitutes for U. The binding density of DNA segment to intron 7nt is determined by the amount of complement sequences.
Intron 7nt was multiplied by 7nt string frequency values of 50kb DNA segments on same row and sum of the products in same column was intron 7nt binding density of this segment DNA (to see following formula and table):

B1*D1+B2*D2+B3*D3+ ······ +B16384*D16384  =E1+E2+E3+ ······ +E16384=  binding density of intron 7nt to DNA segment

The binding density of intron 7nt to DNA segment simulates the total numbers of RNA fragments binding to the DNA segment and is termed RNA binding density.

|  | A | B | C | D | E |
|---|---|---|---|---|---|
|  |  | 7nt string frequency values of DNA segment |  | Intron 7nt | C*D |
| 1 | 5' AAAAAAA 3'<br>3' TTTTTTT 5' | 152 | 5' AAAAAAA 3' | 934.39 | 142027.28 |
| 2 | 5' AAAAAAC 3'<br>3' TTTTTTG 5' | 10 | 5' AAAAAAC 3' | 84.01 | 840.10 |
| 3 | 5' AAAAAAG 3'<br>3' TTTTTTC 5' | 12 | 5' AAAAAAG 3' | 135.89 | 1630.68 |
| …… | …… | …… | …… |  | …… |
| 16384 | 5' TTTTTTT 3'<br>3' AAAAAAA 5' | 186 | 5' TTTTTTT 3' | 1417.7 | 263692.20 |
|  |  |  |  |  | Sum |

The nucleotide sequences of intron 7nt are complement, for example the frequency value of 5'AAAAAAA3' is close to that of 5'TTTTTTT3'. So the binding density of intron 7nt to 5'→3' DNA strand is similar to that to 3'→5' DNA strand. In this paper the results of single strand DNA were reported.

## Results

**The mean values of RNA binding density on 23 chromosomes**

The mean values of RNA binding density on 23 chromosomes are shown in Table 1. The value on *X*-chromosome is significantly lower than that on chromosomes *1,2,3,4,5,6,7,9,10,12,13,14,15,16,17* and *19;* not significantly different from that on chromosomes *8,18,21* and *22*; however higher than that on chromosomes *11* and *20*.

. **Table 1**. The mean values of RNA binding density on 23 chromosomes

| Chromosome | No of segment | Mean±s | vs. X P | Chromosome | No of segment | Mean±s | vs. X P |
|---|---|---|---|---|---|---|---|
| *1* | 4438 | 2562789±279702.6 | <0.001 | *13* | 1915 | 2578853±193533.3 | <0.001 |
| *2* | 4758 | 2555087±228903.3 | <0.001 | *14* | 1745 | 2575179±268075.2 | <0.001 |
| *3* | 3884 | 2564261±223691.9 | <0.001 | *15* | 1632 | 2568135±269287.4 | <0.001 |
| *4* | 3747 | 2577847±192834.1 | <0.001 | *16* | 1602 | 2576405±299706.8 | <0.001 |
| *5* | 3556 | 2557427±226852.5 | <0.001 | *17* | 1559 | 2640048±336271.1 | <0.001 |
| *6* | 3346 | 2569942±207407.6 | <0.001 | *18* | 1495 | 2541916±209847.9 | >0.05 |
| *7* | 3102 | 2598799±257052.9 | <0.001 | *19* | 1117 | 2719852±326802.9 | <0.001 |
| *8* | 2845 | 2538465±222876.4 | >0.05 | *20* | *1192* | *2505988±304984.9* | *<0.001* |
| *9* | 2334 | 2551699±245079.8 | <0.05 | *21* | 682 | 2540904±234180.7 | >0.05 |
| *10* | 2633 | 2550972±272104.3 | <0.05 | *22* | 695 | 2525846±336270.4 | >0.05 |
| *11* | *2620* | *2504660±252672.5* | *<0.001* | X | 2975 | 2536109±212898.1 | |
| *12* | 2595 | 2596026±269705.4 | <0.001 | | | | |

**The amounts of DNA segments highly binding RNAs on *X*-chromosome and autosomes**

The amounts of DNA segments that can highly bind RNAs (intron 7nt) on 23 chromosomes are shown in Table 2. There are 154 statistic analyses (chi-square test) in comparison between *X*-chromosome and 22 autosomes (Table 3). 116 of 154 analyses show that percentages of DNA segments highly binding RNAs on *X*-chromosome are significantly lower than those on autosomes, and 32 analyses not significantly different. Only 6 analyses show that percentages of DNA segments highly binding RNAs on *X*-chromosome are significantly higher than those on autosomes.

The percentages of DNA segments highly binding RNAs on most chromosomes (19/22) are higher than those on *X*-chromosome, and only in some fragments on chromosomes *4,13* and *20* show lower than those on *X*-chromosome (3 for chromosome *4,* 2 for chromosome *13*, 1 for chromosome *20*). If it is considered that on the 3 chromosomes percentages of DNA segments highly binding RNAs are still higher than those on *X*-chromosome and the criterion is used for all comparisons between *X* and autosomes, it is concluded that the percentages of DNA segments highly binding RNAs on *X*-chromosome are significantly lower as compared to all autosomes. While percentages of DNA segments that highly bind RNAs on *X* chromosome are the lowest and on chromosomes *4, 8, 13, 18* and *21* are lower, the percentages on chromosomes *19* is the highest. Figure 1 shows RNA binding density of chromosomes *19* and *X*.

**Table 2.** The amount of DNA segments highly binding RNAs on *X*-chromosome and autosomes

|   | No of segments | RNA binding density | | | | | | |
|---|---|---|---|---|---|---|---|---|
|   |   | >3000000 | >2920000 | >2840000 | >2760000 | >2680000 | >2600000 | >2520000 |
| 1 | 4438 | 316 (7.12) | 449 (10.12) | 634 (14.29) | 863 (19.45) | 1262 (28.44) | 1757 (39.59) | 2426 (54.66) |
| 2 | 4758 | 201 (4.22) | 315 (6.62) | 465 (9.77) | 696 (14.63) | 1123 (23.60) | 1771 (37.22) | 2623 (55.13) |
| 3 | 3884 | 168 (4.33) | 233 (6.00) | 357 (9.19) | 537 (13.83) | 918 (23.64) | 1475 (37.98) | 2262 (58.24) |
| 4 | 3747 | 93 (2.48) | 143 (3.82) | 229 (6.11) | 445 (11.88) | 891 (23.78) | 1592 (42.49) | 2443 (65.20) |
| 5 | 3556 | 145 (4.08) | 193 (5.43) | 290 (8.16) | 485 (13.64) | 806 (22.67) | 1355 (38.10) | 1992 (56.02) |
| 6 | 3346 | 114 (3.41) | 183 (5.47) | 296 (8.85) | 468 (13.99) | 786 (23.49) | 1336 (39.93) | 1992 (59.53) |
| 7 | 3102 | 228 (7.35) | 318 (10.25) | 430 (13.86) | 621 (20.02) | 935 (30.14) | 1408 (45.39) | 1943 (62.64) |
| 8 | 2845 | 80 (2.81) | 132 (4.64) | 220 (7.73) | 339 (11.92) | 585 (20.56) | 998 (35.08) | 1532 (53.85) |
| 9 | 2334 | 104 (4.46) | 164 (7.03) | 245 (10.50) | 368 (15.77) | 566 (24.25) | 879 (37.66) | 1278 (54.76) |
| 10 | 2633 | 164 (6.23) | 243 (9.23) | 346 (13.14) | 491 (18.65) | 695 (26.40) | 1026 (38.97) | 1424 (54.08) |
| 11 | 2620 | 98 (3.74) | 149 (5.69) | 241 (9.20) | 344 (13.13) | 532 (20.31) | 816 (31.15) | 1186 (45.27) |
| 12 | 2595 | 222 (8.55) | 292 (11.25) | 391 (15.07) | 539 (20.77) | 788 (30.37) | 1115 (42.97) | 1546 (59.58) |
| 13 | 1915 | 46 (2.40) | 69 (3.60) | 118 (6.16) | 238 (12.43) | 494 (25.80) | 871 (45.48) | 1254 (65.48) |
| 14 | 1745 | 140 (8.02) | 185 (10.60) | 244 (13.98) | 331 (18.97) | 478 (27.39) | 702 (40.23) | 999 (57.25) |
| 15 | 1632 | 117 (7.17) | 168 (10.29) | 235 (14.40) | 335 (20.53) | 449 (27.51) | 629 (38.54) | 848 (51.96) |
| 16 | 1602 | 155 (9.68) | 210 (13.11) | 283 (17.67) | 362 (22.60) | 491 (30.65) | 663 (41.39) | 892 (55.68) |
| 17 | 1559 | 226 (14.50) | 298 (19.11) | 377 (24.18) | 488 (31.30) | 632 (40.54) | 787 (50.48) | 973 (62.41) |
| 18 | 1495 | 44 (2.94) | 76 (5.08) | 112 (7.49) | 177 (11.84) | 298 (19.93) | 533 (35.65) | 805 (53.85) |
| 19 | 1117 | 232 (20.77) | 323 (28.92) | 419 (37.51) | 524 (46.91) | 605 (54.16) | 697 (62.40) | 778 (69.65) |
| 20 | 1192 | 84 (7.05) | 112 (9.40) | 153 (12.84) | 200 (16.78) | 271 (22.73) | 361 (30.29) | 489 (41.02) |
| 21 | 682 | 19 (2.79) | 31 (4.55) | 48 (7.04) | 89 (13.05) | 167 (24.49) | 289 (42.38) | 400 (58.65) |
| 22 | 695 | 69 (9.93) | 93 (13.38) | 115 (16.55) | 150 (21.58) | 196 (28.20) | 255 (36.69) | 315 (45.32) |
| X | 2975 | 104 (3.50) | 155 (5.21) | 223 (7.50) | 335 (11.26) | 548 (18.42) | 907 (30.49) | 1434 (48.20) |

The numbers in bracket are the percentages of highly binding segments in total segments

**Table 3.** Comparison of the percentages of DNA segments highly binding RNAs between
*X*-chromosome and autosomes

| *X vs.* | Chi-square test | RNA binding density | | | | | | |
|---|---|---|---|---|---|---|---|---|
| | | >3000000 | >2920000 | >2840000 | >2760000 | >2680000 | >2600000 | >2520000 |
| *1* | $X^2$ value | 43.78 | 57.31 | 80.31 | 88.07 | 96.82 | 64.10 | 29.81 |
| | *P* | <0.001 | <0.001 | <0.001 | <0.001 | <0.001 | <0.001 | <0.001 |
| *2* | $X^2$ value | 2.57 | 6.38 | 11.71 | 17.96 | 29.02 | 36.67 | 35.21 |
| | *P* | >0.05 | <0.05 | <0.001 | <0.001 | <0.001 | <0.001 | <0.001 |
| *3* | $X^2$ value | 3.04 | 1.96 | 6.26 | 9.99 | 27.27 | 41.68 | 68.30 |
| | *P* | >0.05 | >0.05 | <0.05 | <0.01 | <0.001 | <0.001 | <0.001 |
| *4* | $X^2$ value | -5.99 | -7.60 | -5.07 | 0.61 | 28.31 | 102.24 | 196.27 |
| | *P* | <0.05 | <0.01 | <0.05 | >0.05 | <0.001 | <0.001 | <0.001 |
| *5* | $X^2$ value | 1.50 | 0.15 | 0.97 | 8.35 | 17.77 | 41.51 | 39.68 |
| | *P* | >0.05 | >0.05 | >0.05 | <0.01 | <0.001 | <0.001 | <0.001 |
| *6* | $X^2$ value | -0.04 | 0.21 | 3.81 | 10.56 | 24.32 | 61.31 | 81.47 |
| | *P* | >0.05 | >0.05 | >0.05 | <0.01 | <0.001 | <0.001 | <0.001 |
| *7* | $X^2$ value | 43.68 | 53.77 | 64.17 | 87.88 | 113.10 | 143.01 | 128.17 |
| | *P* | <0.001 | <0.001 | <0.001 | <0.001 | <0.001 | <0.001 | <0.001 |
| *8* | $X^2$ value | -2.22 | -1.01 | 0.12 | 0.61 | 4.26 | 13.93 | 18.56 |
| | *P* | >0.05 | >0.05 | >0.05 | >0.05 | <0.05 | <0.001 | <0.001 |
| *9* | $X^2$ value | 3.20 | 7.64 | 14.66 | 23.12 | 26.81 | 30.15 | 22.48 |
| | *P* | >0.05 | <0.01 | <0.001 | <0.001 | <0.001 | <0.001 | <0.001 |
| *10* | $X^2$ value | 22.92 | 34.22 | 48.82 | 60.69 | 51.50 | 44.46 | 19.33 |
| | *P* | <0.001 | <0.001 | <0.001 | <0.001 | <0.001 | <0.001 | <0.001 |
| *11* | $X^2$ value | 0.24 | 0.62 | 5.31 | 4.57 | 3.18 | 0.28 | 4.82 |
| | *P* | >0.05 | >0.05 | <0.05 | <0.05 | >0.05 | >0.05 | <0.05 |
| *12* | $X^2$ value | 64.38 | 68.56 | 81.01 | 94.76 | 108.48 | 93.35 | 72.08 |
| | *P* | <0.001 | <0.001 | <0.001 | <0.001 | <0.001 | <0.001 | <0.001 |
| *13* | $X^2$ value | -4.69 | -6.88 | -3.20 | 1.54 | 37.80 | 113.22 | 140.56 |
| | *P* | <0.05 | <0.01 | >0.05 | >0.05 | <0.001 | <0.001 | <0.001 |
| *14* | $X^2$ value | 45.98 | 47.83 | 51.91 | 53.92 | 52.05 | 46.46 | 36.05 |
| | *P* | <0.001 | <0.001 | <0.001 | <0.001 | <0.001 | <0.001 | <0.001 |
| *15* | $X^2$ value | 31.14 | 41.78 | 56.10 | 72.81 | 51.37 | 30.76 | 5.96 |
| | *P* | <0.001 | <0.001 | <0.001 | <0.001 | <0.001 | <0.001 | <0.05 |
| *16* | $X^2$ value | 74.49 | 88.52 | 109.52 | 103.66 | 88.74 | 54.88 | 23.30 |
| | *P* | <0.001 | <0.001 | <0.001 | <0.001 | <0.001 | <0.001 | <0.001 |
| *17* | $X^2$ value | 183.43 | 219.93 | 248.06 | 276.56 | 259.95 | 174.73 | 82.94 |
| | *P* | <0.001 | <0.001 | <0.001 | <0.001 | <0.001 | <0.001 | <0.001 |
| *18* | $X^2$ value | -0.95 | -0.03 | 0.00 | 0.33 | 1.48 | 12.15 | 12.68 |
| | *P* | >0.05 | >0.05 | >0.05 | >0.05 | >0.05 | <0.001 | <0.001 |
| *19* | $X^2$ value | 321.52 | 442.39 | 553.11 | 622.32 | 512.65 | 347.00 | 150.44 |
| | *P* | <0.001 | <0.001 | <0.001 | <0.001 | <0.001 | <0.001 | <0.001 |
| *20* | $X^2$ value | 24.91 | 24.86 | 29.56 | 23.16 | 10.03 | -0.02 | -17.64 |
| | *P* | <0.001 | <0.001 | <0.001 | <0.001 | <0.01 | >0.05 | <0.001 |
| *21* | $X^2$ value | -0.86 | -0.51 | -0.17 | 1.73 | 12.98 | 35.63 | 24.23 |
| | *P* | >0.05 | >0.05 | >0.05 | >0.05 | <0.001 | <0.001 | <0.001 |
| *22* | $X^2$ value | 51.89 | 59.70 | 55.20 | 52.34 | 33.35 | 10.02 | -1.87 |
| | *P* | <0.001 | <0.001 | <0.001 | <0.001 | <0.001 | <0.01 | >0.05 |

## Discussion

This is the first report comparing *X*-chromosome to 22 autosomes for RNA binding density of DNA segments that was used to simulate binding ability of DNA sequences to complement RNA fragments. It was found that the mean value of RNA binding density on *X*-chromosome was significantly lower than that on chromosomes *1,2,3,4,5,6,7,9,10,12,13,14,15,16,17* and *19;* not significantly different from that on chromosomes *8,18,21* and *22*; and higher than that on chromosomes *11* and *20*. The percentages of DNA segments highly binding RNAs were less on *X* than those on all autosomes, although no significant difference (or even higher) was observed in some fragments on some chromosomes.

*X*-chromosome has 2 characteristics: lower mean value of RNA binding density and the lowest percentages of DNA segments highly binding RNAs. If RNAs are same, RNA binding density of a segment of DNA is determined by the sequence composition of the DNA segment. These data indicate that in term of RNA binding density, DNA sequence composition on *X* chromosome is significantly different from that on autosomes. Consistent with this opinion, the assay of gene expression in *X*;autosome translocations (WHITE 1998) and the data of distribution of repetitive elements (LYON 2000) indicate that differences of DNA sequences exist between *X* chromosome and autosomes.

Besides *X*, chromosomes *4, 8, 13, 18* and *21* are those on which percentages of DNA segments that highly bind RNAs are lower. Chromosomes *4,8,13,21* and *X* have been found to have the lowest gene density (Wright *et al.* 2001, Venter *et al* 2001). Chromosomes *4, 13, 18, 21* and *X* show an overall low gene expression and are devoid of RIDGEs (regions of increased gene expression) (Caron et al 2001, Wright *et al.* 2001). Most of chromosome-*18* chromatin represents G-dark bands, and consists mainly of gene-poor and later replicating chromatin (TANABE 2002). It had been noted that the aneuploidies that are compatible with survival until birth (trisomies 13,18 and 21, as well as *X* aneuploidy) appeared to occur in relatively gene-poor chromosomes (Wright *et al.* 2001). These analyses show that low percentages of DNA segments that highly bind RNAs are associated with poor-gene density, low gene expression, survivable aneuploidy cases, G-dark bands, late DNA replication, etc.

Conversely, the percentages of DNA segments that highly bind RNAs on chromosome *19* are the highest among all 23 chromosomes. Chromosome *19* is unusual in many respects such as the high gene density (DEHAL et al 2001), high expression levels (CARON 2001), more transcription-coupled repair sites (SURRALLÉS 2002), mainly early replicating chromatin, and location toward the nuclear center. Most of chromatin on chromosome 19 belongs to G-light bands (TANABE 2002). These findings indicate that high percentages of DNA segments that highly bind RNAs are associated with rich-gene density, high gene expression, G-light bands, early DNA replication, etc. This association is consistent with the analysis results about chromosomes *4, 8, 13, 18, 21* and *X* on which percentages of DNA segments highly binding RNAs are lower.

The levels of percentages of DNA segments that highly bind RNAs are only consistent with the mean values of RNA binding density on some, but not all, chromosomes, for example chromosomes *8,18, 21, X, 17* and *19*. The percentages of DNA segments highly binding RNAs seem to be a more useful marker in order to compare biology of different chromosomes.

*X* chromosomes have particular genomic characteristics, such as the richest L1 repetitive elements (BAILEY et al 2000, LYON 2000), the least GC-rich 50 kb (VENTER 2001) and the least DNA segments that can highly bind RNAs (presented in this study). The *X*-chromosomes appears that one of the two *X* chromosomes of females becomes transcriptionally inactive, which is seen only on *X* chromosomes. What might account for the observed relationship among the least DNA segments that can highly bind RNAs, the richest L1 repetitive elements and XCI?

A lot of ways, such as *cis*-acting elements (WILLARD and CARREL 2001), local gene-specific factors (MCBURNEY 1988), specific chromosomal domains (Miller and Willard 1998), heterogeneous XCI expression (Carrel 1999), and chromatin state (FONG *et al* 2002) seem to be involved to the behaviors of *X* chromosome genes.

To explain the relationship among RNA binding density, XCI, gene density, cytogenetic bands, cis-acting, heterogeneous XCI expression, etc, we propose a model (RNAs opening chromatin) that is conjectural, but is simple and remarkably consistent with many biological facts.

If binding of DNA segments to RNAs opens the chromatin, it is easier for DNA segments on *X* chromosome to become heterochromatin (inactive state) because the numbers of DNA segments highly binding RNAs are significantly less on *X* chromosomes than on autosomes. During mammalian female embryogenesis absence and change of RNAs in cells could result in competition for access to them in 2 *X*-chromosomes, which may be a cause of *X*-chromosome inactivation. *X*-chromosomes satisfy two necessary conditions for competition between two *X*: special DNA sequence structure (different from all autosomes, for example, the richest L1) and the least DNA segments that can highly bind RNAs. Inactive regions replicate later and loose chromatins replicate early (WIDROW et al 1998, GILBERT 2002), which can bring about *cis*-action and influence surrounding regions. In cell cycle the quality and quantity of RNAs change continuously, so the RNAs contacted by early replicating DNA are different from those by later replicating DNA. Therefore early replicating DNA has opportunity to bind more RNAs and maintains sufficiently active state, and vice versa, which may be the reason why the states of the active and inactive *X* chromosomes could be maintained in somatic cells. Although genes are same in different cells, RNAs are variable. Therefore XCI expression may be heterogeneous.

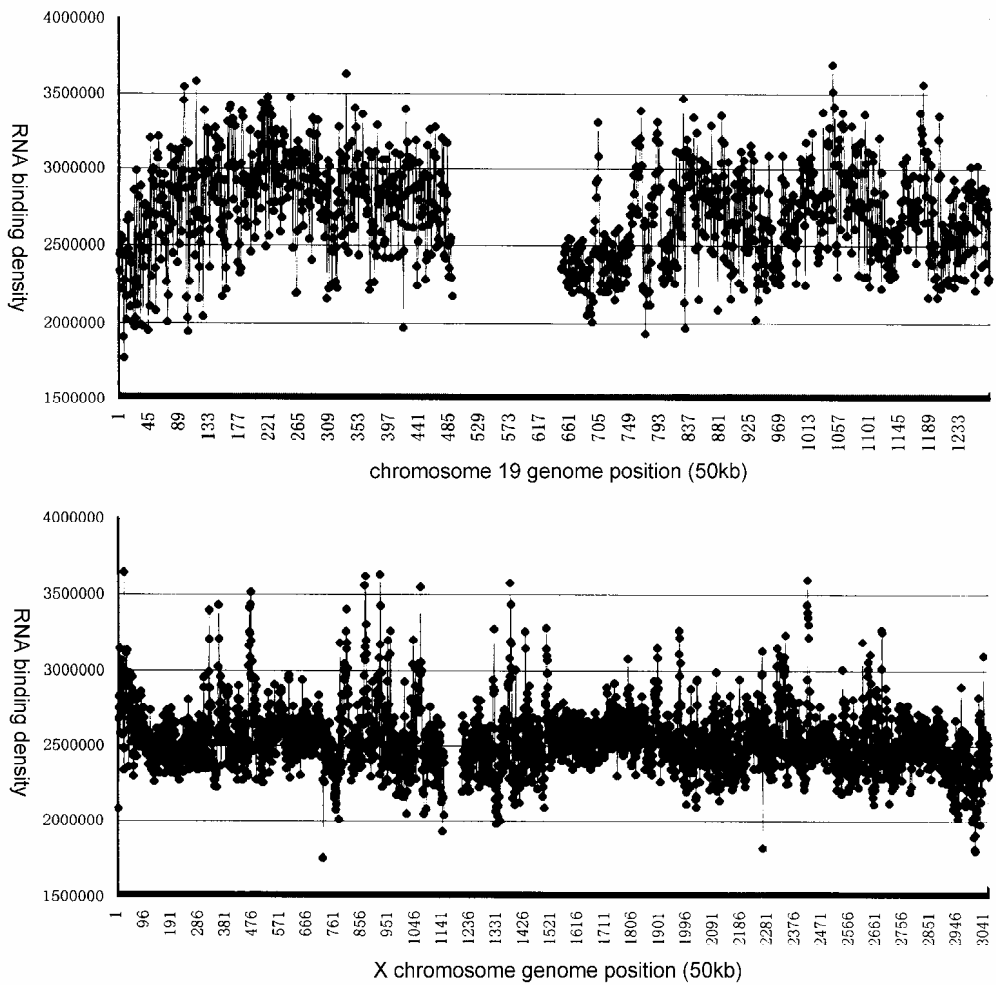

Fig 1. RNA binding density of DNA segments on chromosomes 19 and X